\documentclass{JHEP3}

\usepackage{epsfig}

\def  \bcen   {\begin{center}}
\def  \ecen   {\end{center}}
\def  \beq    {\begin{equation}}
\def  \eeq    {\end{equation}}
\def  \beqa   {\begin{eqnarray}}
\def  \eeqa   {\end{eqnarray}}

\def  \et       {\not\!\! E_T}

\def\bea{\begin{eqnarray}}
\def\eea{\end{eqnarray}}

\def\err#1#2{\lower2pt\hbox{ $\stackrel{\scriptstyle +#1}{\scriptstyle -#2}$}}
\def\ga{\mathrel{\raise.3ex\hbox{$>$\kern-.75em\lower1ex\hbox{$\sim$}}}}
\def\la{\mathrel{\raise.3ex\hbox{$<$\kern-.75em\lower1ex\hbox{$\sim$}}}}
\def\bmaT{\left(\begin{array}{ccc}}
\def\emaT{\end{array}\right)}
\def\bma{\left( \begin{array} }
\def\ema{\end{array} \right)}
\def\gsim{~{\rlap{\lower 3.5pt\hbox{$\mathchar\sim$}}\raise 1pt\hbox{$>$}}\,}
\def\lsim{~{\rlap{\lower 3.5pt\hbox{$\mathchar\sim$}}\raise 1pt\hbox{$<$}}\,}

\title{Leptonic signatures of doubly charged Higgs boson production at the LHC }  
\author{A.~G.~Akeroyd$^{1,2}$, Cheng-Wei Chiang$^{1,3,4}$, Naveen Gaur$^5$ \\
  $^1$ Department of Physics and Center for Mathematics and Theoretical
  Physics \\
  National Central University, Chungli, Taiwan 320, Republic of China. \\
  $^2$ NExT Institute and School of Physics and Astronomy, University of Southampton, \\
  Highfield, Southampton SO17 1BJ, United Kingdom. \\
  $^3$ Institute of Physics, Academia Sinica, Taipei, Taiwan 115, Republic of China. \\
  $^4$ Department of Physics, University of Wisconsin-Madison, Madison, WI
  53706, USA. \\  
  $^5$ Department of Physics, Dyal Singh College (University of Delhi), Lodi Road \\
  New Delhi - 110003, India. \\
  E-mail: \email{a.g.akeroyd, chengwei@phys.ncu.edu.tw, gaur.nav@gmail.com},
}

\preprint{SHEP-10-29, \\ arXiv:1009.2780 [hep-ph]}

\abstract{
The production of doubly charged Higgs bosons ($H^{\pm\pm}$) at the
CERN LHC can give rise to distinctive multi-lepton signatures.  The
discovery potential of $H^{\pm\pm}$ can be optimized by considering a
search strategy which is sensitive to both of the dominant production
mechanisms, $q\overline q\to H^{++}H^{--}$ and $q\overline {q'}\to
H^{\pm\pm}H^{\mp}$. We compare the discovery potential for the
signatures of exactly four leptons and at least three leptons in the
final state, using the same set of cuts. We have carried out fast
detector simulations at the LHC for both signal and backgrounds for a
wide range of values of the charged Higgs mass. We find that the use
of the latter channel can substantially improve the detection
prospects of the doubly charged Higgs boson at the LHC. 
}

\keywords{Doubly charged Higgs boson, Higgs Triplet Model} 

\begin{document}

\newcommand{\gev}{\, {\rm GeV}}
\newcommand{\MNS}{{\textrm{MNS}}}
\newcommand{\eV}{\textrm{eV}}
\newcommand{\GeV}{\textrm{GeV}}
\newcommand{\TeV}{\textrm{TeV}}
\newcommand{\Hlilj}{H^{\pm\pm}\to \ell_i^\pm \ell_j^\pm}
\newcommand{\Hlili}{H^{\pm\pm}\to \ell_i^\pm \ell_i^\pm}
\newcommand{\BR}{\textrm{BR}}
\newcommand{\BRll}{\textrm{BR}( H^{\pm\pm} \to \ell^\pm \ell^\pm )}
\newcommand{\BRlili}{\textrm{BR}( H^{\pm\pm} \to \ell_i^\pm \ell_i^\pm )}
\newcommand{\BRlilj}{\textrm{BR}( H^{\pm\pm} \to \ell_i^\pm \ell_j^\pm )}
\newcommand{\BRee}{\textrm{BR}( H^{\pm\pm} \to e^\pm e^\pm )}
\newcommand{\BRem}{\textrm{BR}( H^{\pm\pm} \to e^\pm \mu^\pm )}
\newcommand{\BRet}{\textrm{BR}( H^{\pm\pm} \to e^\pm \tau^\pm )}
\newcommand{\BRmm}{\textrm{BR}( H^{\pm\pm} \to \mu^\pm \mu^\pm )}
\newcommand{\BRmt}{\textrm{BR}( H^{\pm\pm} \to \mu^\pm \tau^\pm )}
\newcommand{\BRtt}{\textrm{BR}( H^{\pm\pm} \to \tau^\pm \tau^\pm )}
\newcommand{\rel}{{\textrm{rel}}}
\newcommand{\true}{{\textrm{true}}}
\newcommand{\fit}{{\textrm{fit}}}
\newcommand{\theory}{{\textrm{theory}}}
\newcommand{\eff}{{\textrm{eff}}}
\newcommand{\BG}{{\textrm{BG}}}
\newcommand{\CPC}{{\textrm{CPC case}}}
\newcommand{\CPV}{{\textrm{CPV}}}
\newcommand{\allow}{{\textrm{allow}}}
\newcommand{\hierarchy}{\textrm{hierarchy}}
\newcommand{\magic}{{\textrm{mgc}}}
\newcommand{\sweet}{{\textrm{ss}}}
\newcommand{\tlll}{{\tau \to \bar{l}_i l_j l_k}}
\newcommand{\tmmm}{{\tau \to \bar{\mu}\mu\mu}}
\newcommand{\tmme}{{\tau \to \bar{\mu}\mu e}}
\newcommand{\temm}{{\tau \to \bar{e}\mu\mu}}
\newcommand{\tmee}{{\tau \to \bar{\mu}ee}}
\newcommand{\teme}{{\tau \to \bar{e}\mu e}}
\newcommand{\meee}{{\mu \to \bar{e}ee}}
\newcommand{\meg}{{\mu \to e\gamma}}
\newcommand{\tr}{{\rm Tr}}
\newcommand{\ratio}[2]{\textrm{BR($#1$)/BR($#2$)}}
\newcommand{\calchep}{CalcHEP }
\newcommand{\pythia}{{\sc pythia }}
\newcommand{\atlfast}{{\sc atlfast }}

\maketitle



\section{Introduction \label{section:1}}

The firm evidence that neutrinos oscillate and possess small masses
below the eV scale~\cite{Fukuda:1998mi} necessitates physics beyond
the Standard Model (SM), which could manifest itself at the CERN Large
Hadron Collider (LHC) and/or in low energy experiments which search
for lepton flavour violation (LFV)~\cite{Kuno:1999jp}.  Consequently,
models of neutrino mass generation which can be probed at present and
forthcoming experiments are of great phenomenological interest.  

Neutrinos may obtain masses via the vacuum expectation value (vev) of
a neutral Higgs boson in an isospin triplet
representation~\cite{Konetschny:1977bn, Mohapatra:1979ia, Magg:1980ut,
Schechter:1980gr, Cheng:1980qt}.  A particularly simple implementation
of this mechanism of neutrino mass generation is the ``Higgs Triplet
Model'' (HTM) in which the SM Lagrangian is augmented solely by an
$SU(2)$ triplet of scalar particles with hypercharge
$Y=2$~\cite{Konetschny:1977bn,Schechter:1980gr,Cheng:1980qt}.  In the
HTM, neutrinos acquire Majorana masses given by the product of a
triplet Yukawa coupling ($h_{ij}$) and a triplet vev ($v_\Delta$).
Consequently, there is a direct connection between $h_{ij}$ and the
neutrino mass matrix, which gives rise to phenomenological predictions
for processes which depend on $h_{ij}$
\cite{Ma:1998dx,Chun:2003ej,Kakizaki:2003jk,Garayoa:2007fw,Akeroyd:2007zv,Kadastik:2007yd,Perez:2008ha,delAguila:2008cj,Petcov:2009zr,Akeroyd:2009nu,Akeroyd:2009hb,Fukuyama:2009xk}.
A distinctive signal of the HTM would be the observation of a doubly
charged Higgs boson ($H^{\pm\pm}$), whose mass ($M_{H^{\pm\pm}}$) may
be of the order of the electroweak scale. Such particles can be
produced with sizeable rates at hadron colliders in the processes
$q\overline q\to H^{++}H^{--}$
\cite{Barger:1982cy,Gunion:1989in,Muhlleitner:2003me,Huitu:1996su,Han:2007bk}
and $q\overline {q'}\to
H^{\pm\pm}H^{\mp}$~\cite{Barger:1982cy,Dion:1998pw, Akeroyd:2005gt},
where $H^\pm$ is a singly charged Higgs boson in the same triplet
representation.  Direct searches for $H^{\pm\pm}$ have been carried
out at the Fermilab Tevatron, assuming the production channel
$q\overline q\to H^{++}H^{--}$ and {\it the leptonic} decays
$H^{\pm\pm}\to \ell^\pm_i\ell^\pm_j$ ($\ell=e,\mu,\tau$), and mass
limits in the range $M_{H^{\pm\pm}}> 110 - 150$~GeV have been obtained
\cite{Acosta:2004uj,Abazov:2004au,Abazov:2008iy,Aaltonen:2008ip}.  The
CERN Large Hadron Collider (LHC), using the above production
mechanisms, will offer improved sensitivity to $M_{H^{\pm\pm}}$
\cite{Azuelos:2005uc,Han:2007bk,Rommerskirchen:2007jv,Hektor:2007uu,Perez:2008ha,delAguila:2008cj}.

The decay channels $H^{\pm\pm}\to \ell^\pm\ell^\pm$ and
$H^{\pm}\to \ell^\pm\nu$ are the dominant ones if $v_\Delta\lsim
10^{-4}$ GeV, and give rise to multi-lepton signatures. In the HTM,
one expects $v_\Delta\lsim 10^{-4}$ GeV if the triplet Yukawa coupling
is larger than the smallest Yukawa coupling in the SM ({\it i.e.}, the
electron Yukawa coupling).  One can define various multi-lepton
signatures which originate from the production mechanisms $q\overline
q\to H^{++}H^{--}$ and $q\overline {q'}\to H^{\pm\pm}H^{\mp}$.  The
four-lepton signature ($4\ell$) only receives a contribution from
$q\overline q\to H^{++}H^{--}$, and the detection prospects in this
channel at the LHC have been studied in \cite{Azuelos:2005uc,
  Han:2007bk,Rommerskirchen:2007jv,Hektor:2007uu,delAguila:2008cj}.
Although this $4\ell$ signature provides a very promising way to
search for $H^{\pm\pm}$, it is not necessarily the channel which
offers the best sensitivity for a given integrated luminosity and mass
$M_{H^{\pm\pm}}$. Recently, attention has been given to the
three-lepton channel \cite{Perez:2008ha,delAguila:2008cj}, which also
has relatively small SM backgrounds. Importantly, the signature of
three-leptons is sensitive to the production mechanism $q\overline
{q'}\to H^{\pm\pm}H^{\mp}$ \cite{Akeroyd:2005gt}. The magnitudes of
the cross sections of $q\overline q\to H^{++}H^{--}$ and $q\overline
{q'}\to H^{\pm\pm}H^{\mp}$ are comparable in a large parameter space
of the HTM, because the scalar potential of the model gives
$M_{H^{\pm\pm}}\sim M_{H^{\pm}}$ (unless a specific scalar quartic
coupling is taken to be {\it fairly} large). In order to improve the
sensitivity of the LHC to $M_{H^{\pm\pm}}$, one can define two
distinct signatures consisting of three leptons, in which two of the
leptons have the same electric charge and are assumed to originate
from $H^{\pm\pm}$ (which is produced in both production mechanisms,
$q\overline q\to H^{++}H^{--}$ and $q\overline {q'}\to
H^{\pm\pm}H^{\mp}$): i) the signature of ``exactly three leptons''
($3\ell$), \cite{delAguila:2008cj}, and ii) the signature of ``three
or more leptons'' ($\ge 3\ell$) \cite{Abazov:2008iy}. Different
sensitivity to $M_{H^{\pm\pm}}$ is expected in these two channels. 

Detection prospects at the LHC are best for $\ell=e,\mu$, for which
there are several exclusive three-lepton channels, {\it e.g.}
$eee,\mu\mu\mu,e\mu\mu$ etc \cite{Akeroyd:2009hb}. However, in the
region of high invariant mass for a pair of same-sign leptons ({\it
e.g.}, $m_{\ell\ell}> 200$ GeV) one expects similar detection
efficiencies and SM backgrounds in these exclusive three-lepton
channels with $\ell=e,\mu$. Therefore, to optimise the sensitivity to
$M_{H^{\pm\pm}}$ it is reasonable to define an inclusive signature in
which $e^\pm$ and $\mu^\pm$ are treated as identical particles, and
this is the approach which is taken in \cite{delAguila:2008cj}. The
signature of $3\ell$ (with $\ell=e,\mu$) is studied in
\cite{delAguila:2008cj} and it is shown that the sensitivity to
$M_{H^{\pm\pm}}$ is significantly superior to that of the $4\ell$
channel.  This very promising result can be further improved, because
the $\ge 3\ell$ channel is expected to give even greater sensitivity
to $M_{H^{\pm\pm}}$ than the $3\ell$ channel. 
In this work we study the signature of $\ge 3\ell$ and compare its
sensitivity to $M_{H^{\pm\pm}}$ with that obtained for the $4\ell$
channel. Our study is the first simulation of the signature $\ge
3\ell$ at LHC that includes both production mechanisms $q\overline
q\to H^{++}H^{--}$ and $q\overline {q'}\to H^{\pm\pm}H^{\mp}$.  We
show that the $\ge 3\ell$ channel is the optimum search strategy for
$H^{\pm\pm}$.  We note that the most recent search by the D0
collaboration \cite{Abazov:2008iy} uses the strategy of $>3\ell$ in
the context of a search for three muons ($\mu^\pm\mu^\pm\mu^\mp$),
with the assumption that production of $H^{\pm\pm}$ is only by $q
\overline{q} \to H^{++} H^{--}$. 

The paper is organized as follows.  In section~\ref{section:2} we will
briefly review the HTM model.  In section~\ref{section:3} we discuss
the production of doubly charged Higgs bosons at hadronic colliders.
In the same section we will describe our analysis setup and framework
for simulations of backgrounds and signal processes.  The results of
our simulations are given in section~\ref{section:4}.  Finally we will
conclude with a summary of our results in section~\ref{section:5}


\section{The Higgs Triplet Model \label{section:2} }

The HTM model \cite{Konetschny:1977bn, Schechter:1980gr, Cheng:1980qt}
 is an extension of the SM in which only the scalar sector is
 augmented with a Higgs triplet. It is a particularly simple model
which contains a doubly charged scalar. The model has the following
$SU(2)\otimes U(1)_Y$ gauge-invariant Yukawa interactions: 
\begin{equation}
{\cal L} \ni h_{ij}\psi_{iL}^TCi\sigma_2\Delta\psi_{jL} + \mbox{h.c.} ~,
\label{trip_yuk}
\end{equation}
where the triplet Yukawa couplings $h_{ij} (i,j=e,\mu,\tau)$ are
complex and symmetric, $C$ is the Dirac charge conjugation operator,
$\sigma_2$ is a Pauli matrix, $\psi_{iL}=(\nu_i, l_i)_L^T$ is a
left-handed lepton doublet, and $\Delta$ is a $2\times 2$
representation of the $Y=2$ complex triplet fields
($\delta^{++},\delta^+,\delta^0$): 
\begin{equation}
\Delta
=\bma{cc}
\delta^+/\sqrt{2}  & \delta^{++} \\
\delta^0       & -\delta^+/\sqrt{2}
\ema ~.
\end{equation}
Note that the mass eigenstate $H^{\pm\pm}$ is entirely composed of the
triplet field ($H^{\pm\pm}\equiv \delta^{\pm\pm}$), while $H^\pm$ is
predominantly $\delta^\pm$, with a small component of isospin doublet
scalar ($\Phi$).  A non-zero Higgs triplet VEV, $\langle
\delta^0\rangle = v_\Delta / \sqrt{2}$, gives rise to the following
Majorana mass matrix for neutrinos: 
\begin{equation}
m_{ij}=2h_{ij}\langle \delta^0\rangle = \sqrt{2}h_{ij}v_{\Delta} ~.
\label{nu_mass}
\end{equation}

Realistic neutrino masses can be obtained with a perturbative $h_{ij}$
provided that $v_{\Delta}\gsim 1$ eV.  The presence of a non-zero
$v_{\Delta}$ gives rise to $\rho\ne 1$ at tree level, where
$\rho\equiv M^2_W/(M^2_Z \cos^2\theta_W$). Therefore $v_{\Delta} \lsim
1$ GeV is necessary in order to comply with the measurement of
$\rho\sim 1$. This simple expression of tree-level masses for the
observed neutrinos is essentially the main motivation for studying the
HTM.  It provides a direct connection between $h_{ij}$ and the
neutrino mass matrix, which gives rise to phenomenological predictions
for processes which depend on $h_{ij}$
\cite{Ma:1998dx,Chun:2003ej,Kakizaki:2003jk,Garayoa:2007fw,Akeroyd:2007zv,%
Kadastik:2007yd,Perez:2008ha,delAguila:2008cj,Petcov:2009zr,Akeroyd:2009nu,%
Akeroyd:2009hb,Fukuyama:2009xk}.

The mass matrix $m_{ij}$ for three Dirac neutrinos is diagonalized by
the PMNS (Pontecorvo-Maki-Nakagawa-Sakata) matrix $V_{\rm
PMNS}$~\cite{Pontecorvo:1957qd}.  For Majorana neutrinos (which is the
case in HTM), two additional phases appear, and then the mixing matrix
$V$ becomes 
\begin{eqnarray}
 V = V_{\rm PMNS} \times
     \textrm{diag}( 1, e^{i\phi_1 /2}, e^{i\phi_2 /2}),
\end{eqnarray}
where $\phi_1$ and $\phi_2$ are referred to as the Majorana
phases~\cite{Schechter:1980gr,Bilenky:1980cx} and $-\pi \le
\phi_1,\phi_2 < \pi$.  One has the freedom to work in the basis in
which the charged lepton mass matrix is diagonal, and then the
neutrino mass matrix is diagonalized by $V_{\rm PMNS}$.  Using
Eq.~(\ref{nu_mass}) one can write the couplings $h_{ij}$ as
follows~\cite{Ma:1998dx,Chun:2003ej}: 
\begin{equation}
h_{ij}
= \frac{m_{ij}}{\sqrt{2}v_\Delta}
\equiv \frac{1}{\sqrt{2}v_\Delta}
\left[
 V_{\rm PMNS}
 \textrm{diag}(m_1,m_2 e^{i\phi_1},m_3 e^{i\phi_2})
 V_{\rm PMNS}^T
\right]_{ij} ~.
\label{hij}
\end{equation}
Here $m_1,m_2$ and $m_3$ are the absolute masses of the three
neutrinos.  Neutrino oscillation experiments are sensitive to
mass-squared differences, $\Delta m_{21}^2$($\equiv m^2_2-m^2_1)$ and
$\Delta m_{31}^2$($\equiv m^2_3-m^2_1$).  Since the sign of $\Delta
m_{31}^2$ is undetermined at present, distinct patterns for the
neutrino mass hierarchy are possible.  The case with $\Delta m^2_{31}
>0$ is referred to as {\it normal hierarchy} (NH) where $m_1 < m_2 <
m_3$, and the case with $\Delta m^2_{31} <0$ is known as {\it inverted
hierarchy} (IH) where $m_3 < m_1 < m_2$.  However, information on the
mass $m_0$ of the lightest neutrino (either $m_1$ or $m_3$) and the
Majorana phases cannot be obtained from neutrino oscillation
experiments.  This is because the oscillation probabilities are
independent of these parameters, not only in vacuum but also in
matter.  An attractive feature of the HTM is this simple relationship
between the triplet Yukawa couplings and the parameters of the
neutrino mass matrix (many of which are measurable) given in
Eq.~(\ref{hij}).  In contrast, Eq.~(\ref{hij}) does not hold in other
models with a doubly charged scalar ({\it e.g.}, the Left-Right
symmetric model in which the couplings $h_{ij}$ are essentially
arbitrary.) 

In this work we are concerned with the case of the leptonic decays of
$H^{\pm\pm}$ dominating, which is realized if $h_{ij}$ are larger than
the smallest Yukawa coupling in the SM ({\it i.e.,} the electron
Yukawa coupling, $h_e \sim 10^{-6}$).  From Eq.~(\ref{hij}) it follows
that $v_\Delta\lsim 0.1$ MeV in this scenario, and thus the decay
$H^{\pm\pm}\to W^\pm W^\pm$ (which depends on $v_\Delta$) is
negligible. Moreover, in the HTM the lifetime of $H^{\pm\pm}$ is
always short enough to ensure that it decays in the detectors of the
Tevatron and the LHC \cite{Han:2007bk}. For very small values of
$v_\Delta$ ({\it e.g.,} $v_\Delta < 10$ eV), the magnitude of the
Yukawa couplings $h_{ij}$ approaches unity. In this case, the charged
scalars (which we assume to have a mass of the order of the
electroweak scale) would induce potentially observable BR's for LFV
decays such as $\mu\to e\gamma$, $\mu\to eee$ and $\tau\to lll$ ({\it
e.g.}, see \cite{Cuypers:1996ia}), whose decay rates depend on the
parameters of the neutrino mass matrix, the absolute values of
$h_{ij}$, and $M_{H^{\pm\pm}}$ or $M_{H^\pm}$
\cite{Chun:2003ej,Kakizaki:2003jk,Akeroyd:2009nu}. Such constraints
can be satisfied for appropriately small $h_{ij}$, the most severe
constraint being from $\mu\to eee$ (which gives $h_{ee}h_{e\mu}\lsim
10^{-7}$). In the parameter space $10^{-6} \lsim h_{ij} \lsim
10^{-3}$, the constraints from the above LFV decays are satisfied,
even for values of $M_{H^{\pm\pm}}$ of the order of the electroweak
scale. The BR of $H^{\pm\pm}\to \ell^\pm \ell^\pm$ depends on the six
parameters of the neutrino mixing matrix, $V$, (with the dominant
uncertainty arising from the unknown Majorana phases, $\phi_1$ and
$\phi_2$), the unknown mass of the lightest neutrino ($m_0$), the mass
splittings of the neutrinos, and the ignorance of the neutrino mass
hierarchy (normal or inverted) \cite{Chun:2003ej}. 
	
Detailed studies of BR$(H^{\pm\pm}\to \ell^\pm \ell^\pm)$ have been
performed in
\cite{Garayoa:2007fw,Akeroyd:2007zv,Kadastik:2007yd,Perez:2008ha}.
Notably, BR$(H^\pm\to \ell^\pm \nu)$ (in which the three flavours of
neutrinos are summed over) does not depend on the Majorana phases, and
the dominant uncertainty is from $m_0$ and the neutrino mass hierarchy
\cite{Perez:2008ha}.  Importantly, BR$(H^{\pm\pm}\to \ell^\pm
\ell^\pm)\sim 100\%$ and BR$(H^\pm\to \ell^\pm \nu)\sim 100\%$ for a
given lepton flavour are not possible in the HTM. Moreover,
BR$(H^{\pm\pm}\to \ell^\pm \ell^\pm)\sim 100\%$ and BR$(H^\pm\to
\ell^\pm \nu)\sim 100\%$ are not possible even when summing over
$\ell=e,\mu$, although values as high as $\sim 70\%$ are possible in
specific regions of the parameter space of the neutrino mass matrix.
A generic model containing $H^{\pm\pm}$ can have BR$(H^{\pm\pm}\to
\ell^\pm \ell^\pm)\sim 100\%$ (summing over $\ell=e,\mu$) for
appropriately chosen $h_{ij}$, {\it e.g.}, the LR symmetric model with
$h_{ee},h_{e\mu},h_{\mu\mu} \gg h_{e\tau},h_{\mu\tau},h_{\tau\tau}$.
In direct searches for $H^{\pm\pm}$ the derived lower limits on
$M_{H^{\pm\pm}}$ are usually given for the extreme case of BR=$100\%$,
and this will be discussed in more detail in the next section. 


\section{Searches for doubly charged Higgs bosons at hadron colliders
\label{section:3}} 

Direct searches for $H^{\pm\pm}$ have been carried out at the Fermilab
Tevatron
\cite{Acosta:2004uj,Abazov:2004au,Abazov:2008iy,Aaltonen:2008ip},
assuming the production mechanism at partonic level given by
$q\overline q \to \gamma^*, Z^* \to H^{++}H^{--}$, whose cross section
depends on only one unknown parameter, $M_{H^{\pm\pm}}$. Production
mechanisms which depend on the triplet VEV ($q\overline {q'}\to W^{\pm
*}\to W^\mp H^{\pm\pm}$ and fusion via $W^{\pm *} W^{\pm *} \to
H^{\pm\pm}$ \cite{Huitu:1996su,Vega:1989tt}) are not competitive with
$q\overline q \to H^{++}H^{--}$ at the energies of the Tevatron. All
searches assume the leptonic decay mode $H^{\pm\pm}\to \ell^\pm
\ell^\pm$, for which there are six possibilities
($ee,\mu\mu,\tau\tau,e\mu,e\tau,\mu\tau$). 
 
The CDF collaboration searched for three final states, $H^{\pm\pm}\to
e^\pm e^\pm, e^\pm \mu^\pm, \mu^\pm\mu^\pm$, requiring at least one
pair of same-sign leptons with high invariant mass
\cite{Acosta:2004uj}.  The integrated luminosity used was $0.24$
fb$^{-1}$ and the mass limits $M_{H^{\pm\pm}}>133,113,136$ GeV were
obtained for the decay channels $H^{\pm\pm}\to e^\pm e^\pm,e^\pm
\mu^\pm, \mu^\pm\mu^\pm$, respectively, assuming BR=$100\%$ in a given
channel.  The D0 collaboration \cite{Abazov:2004au,Abazov:2008iy}
searched for $H^{\pm\pm}\to \mu^\pm\mu^\pm$, and derived the mass
limit $M_{H^{\pm\pm}}> 150$ GeV \cite{Abazov:2008iy} using $1.1$
fb$^{-1}$ of integrated luminosity. The main difference between these
searches by D0 is the requirement in the most recent search
\cite{Abazov:2008iy} of a third $\mu$ of opposite sign to the two
same-sign $\mu$, where the latter is assumed to originate from the
decay of one of the pair-produced $H^{\pm\pm}$. This extra requirement
suppresses backgrounds from $\gamma/Z\to \mu^+\mu^-$ and multijets,
which were less than one event for the integrated luminosity of $0.11$
fb$^{-1}$ used in \cite{Abazov:2004au}, but became non-negligible for
the search in \cite{Abazov:2008iy} with $1.1$ fb$^{-1}$.  The
requirement of a third lepton is necessary for the future Tevatron
searches in order to reduce the SM backgrounds. At the Tevatron, the
main backgrounds to the three-lepton signal with $\ell=e,\mu$ are from
$WZ$ and $ZZ$ production. Two decay channels involving $\tau$
($H^{\pm\pm}\to e^\pm \tau^\pm, \mu^\pm \tau^\pm$), were searched for
by the CDF collaboration in \cite{Aaltonen:2008ip}, and there has been
no search for $H^{\pm\pm}\to \tau^\pm \tau^\pm$. 

As discussed in Section \ref{section:2}, in the HTM one has
BR$(H^{\pm\pm}\to \ell^\pm \ell^\pm)<100\%$ in a given channel, and
thus the above mass limits for $M_{H^{\pm\pm}}$ (which assume
BR=$100\%$) are weakened when applied to the HTM. The expected number
of $H^{\pm\pm}\to \ell^\pm \ell^\pm$ events scales linearly in BR (for
searches for a single pair of same-sign leptons
\cite{Acosta:2004uj,Abazov:2004au}) or quadratically in BR (for
searches which require a third lepton or more \cite{Abazov:2008iy}),
and so the mass limits are weakened accordingly. 

All the above searches at the Tevatron assume {\it only} the
production mechanism $q\overline q\to \gamma^*,Z^*\to H^{++}H^{--}$.
However, the partonic process $q\overline {q'}\to W^*\to
H^{\pm\pm}H^\mp$ \cite{Barger:1982cy,Dion:1998pw,Akeroyd:2005gt} has a
cross section at hadron colliders comparable to that of $q\overline
q\to H^{++}H^{--}$ for $M_{H^\pm}\sim M_{H^{\pm\pm}}$, and thus the
former will also contribute to the search for $H^{\pm\pm}$.  In
Ref.~\cite{Akeroyd:2005gt}, it is suggested that the search potential
at hadron colliders can be improved by considering the following
inclusive single $H^{\pm\pm}$ cross section ($\sigma_{H^{\pm\pm}}$): 
%
%
\begin{equation}
\sigma_{H^{\pm\pm}}=\sigma(p\overline p,pp\to H^{++}H^{--})+
\sigma(p\overline p,pp\to H^{++} H^-)+\sigma(p\overline p,pp \to 
H^{--} H^+)
\label{single_prod}
\end{equation}
At the Tevatron $\sigma(p\overline p\to H^{++} H^-)
=\sigma(p\overline p\to 
H^{--} H^+)$ while at the LHC 
$\sigma(pp\to H^{++} H^-)> \sigma(pp\to 
H^{--} H^+)$.
These two production mechanisms
have different QCD $K$ factors. Explicit calculations
\cite{Muhlleitner:2003me} for $p\overline p,pp\to H^{++}H^{--}$ give
around $K=1.3$ at the Tevatron and $K=1.25$ at the LHC, with a
dependence on $M_{H^{\pm\pm}}$.  In reality, the $K$ factor for
$p\overline p,pp\to H^{\pm\pm}H^{\mp}$ is expected to be very similar
(but not identical) to that for $p\overline p,pp\to H^{++}H^{--}$,
with some dependence on the mass splitting
$M_{H^{\pm\pm}}-M_{H^{\pm}}$. In \cite{Akeroyd:2005gt,Perez:2008ha}
the $K$ factors are taken to be equal. We note that $p\overline
p,pp\to H^{++}H^{--}$ also receives a contribution from real photon
annihilation \cite{Han:2007bk}, which causes an increase in the cross
section of around $10\%$ at the LHC, but much less at the Tevatron. In
our simulation analysis however we do not include this correction.


\section{Simulations of signal and backgrounds at the LHC
\label{section:4}} 

Several studies have been performed to study the doubly charged Higgs
in the decay channel $\Hlilj$ ($i, j=e,\mu,\tau$) at the LHC. The
production mechanism $q\overline q\to \gamma^*,Z^*\to H^{++}H^{--}$
followed by decay $H^{++}H^{--}\to \ell^+\ell^+\ell^-\ell^-$ is
studied in \cite{Perez:2008ha, Han:2007bk, Azuelos:2005uc,
Rommerskirchen:2007jv, Hektor:2007uu, delAguila:2008cj}.  Only two
among these studies also take into account the production mechanism
$pp\to W^{\pm *}\to H^{\pm\pm}H^\mp$
\cite{Perez:2008ha,delAguila:2008cj}, followed by the decays $\Hlilj$
($i, j=e,\mu,\tau$) and $H^\pm\to \ell^\pm_i\nu$.  The LHC sensitivity
to $\Hlilj$ considerably extends that at the Tevatron, due to the
increased cross sections and larger luminosities.  The analysis of
Ref.~\cite{Rommerskirchen:2007jv} shows that $H^{\pm\pm}$ can be
discovered for $m_{H^{\pm\pm}}< 800$~GeV and ${\cal L}=50$~fb$^{-1}$,
assuming $\BRmm =100\%$.  Importantly, all the above simulations
suggest that as little as ${\cal L}= 1~$fb$^{-1}$ is needed for the
discovery of $m_{H^{\pm\pm}}<400$~GeV if one of BR($H^{\pm\pm}\to
e^\pm e^\pm, e^\pm \mu^\pm, \mu^\pm \mu^\pm$) is large, and hence such
a light $H^{\pm\pm}$ would be found very quickly at the LHC.  The
signal from $q\overline q\to \gamma^*,Z^*\to H^{++}H^{--}$ and decay
$H^{++}H^{--}\to \ell^+\ell^+\ell^-\ell^-$ is usually taken to be four
leptons, which are isolated and have sufficiently large transverse
energy.  In Ref.~\cite{Rommerskirchen:2007jv} the signal is taken to
be $4\mu$.  In Ref.~\cite{Hektor:2007uu}, where little Higgs models
are considered, the signal is defined as $4\ell$ where $\ell=\mu,\tau$
(and $e$ is not included), and five different four-lepton signatures
are studied (one of which being $4\mu$).  In Ref.~\cite{Han:2007bk}
$e$ and $\mu$ are treated as the same particle, and a parton-level
study of the four-lepton signature is performed.  In
Ref.~\cite{Azuelos:2005uc} two signatures are defined: i) four leptons
and ii) at least three leptons. It is shown that superior sensitivity
to $M_{H^{\pm\pm}}$ is obtained for the signature of at least three
leptons. 

As discussed earlier, the production mechanism $pp\to W^{\pm *}\to
H^{\pm\pm}H^\mp$ will contribute to the signal for $H^{\pm\pm}$ if
three (or more) leptons are required.  The simulation in
Ref.~\cite{delAguila:2008cj} is the first study of the mechanism
$pp\to H^{\pm\pm}H^\mp$ together with $pp\to H^{++}H^{--}$, with the
aim of improving the sensitivity to $M_{H^{\pm\pm}}$ at the LHC.  In
Ref.~\cite{delAguila:2008cj} $e$ and $\mu$ are not distinguished, and
such an inclusive channel has the advantage of maximizing the
sensitivity to $M_{H^{\pm\pm}}$ for a given integrated luminosity, and
for the general case of BR $< 100\%$ for a given flavour of lepton.
Both a four-lepton signature and a three-lepton signature are studied,
and the sensitivity to $M_{H^{\pm\pm}}$ for the two signatures is
compared, assuming $M_{H^{\pm\pm}}=M_{H^{\pm}}$.  The three-lepton
signature is defined as being {\it exactly} three leptons ($3\ell$),
{\it i.e.}, a fourth lepton is vetoed.  Note that this three-lepton
signature differs from that defined in the latest search for
$H^{\pm\pm}$ at the Tevatron \cite{Abazov:2008iy} in which a fourth
lepton is not vetoed ($\ge 3\ell$). 

In Ref.~\cite{delAguila:2008cj} it is concluded that the three-lepton
signature offers considerably greater discovery potential for $H^{\pm
\pm}$ in the HTM than the signature of four leptons (note that the
same conclusion is obtained in \cite{Azuelos:2005uc}, even without
including $pp\to H^{\pm\pm}H^\mp$).  The main reason for the superior
sensitivity of the three-lepton signature in \cite{delAguila:2008cj}
is the extra contribution from $pp\to H^{\pm\pm}H^\mp$ (which does not
contribute to the four-lepton signature).  Although the SM background
for the three-lepton signature is larger than that for the four-lepton
signature, in the region of high invariant mass of $\ell^\pm\ell^\pm$
(relevant for $M_{H^{\pm\pm}}>200$ GeV) the backgrounds are still
sufficiently small, which gives rise to superior sensitivity to
$M_{H^{\pm\pm}}$ for the three-lepton signature. 
Moreover, Ref.~\cite{delAguila:2008cj} used different sets of cuts
for the three-lepton and four-lepton signatures.

In Ref.~\cite{Perez:2008ha}, a parton-level study at the LHC was
performed for the detection prospects of the production channel $pp\to
H^{\pm\pm}H^\mp$ alone, followed by the decays $H^{\pm\pm}\to \ell^\pm
\ell^\pm$ and $H^\pm\to \ell^\pm\nu$, where both $e$ and $\mu$
contributions are summed together in an inclusive approach like that
in Ref.~\cite{delAguila:2008cj}.  The strategy in \cite{Perez:2008ha}
is to isolate the contribution from $pp\to H^{\pm\pm}H^\mp$ and remove
that from $pp\to H^{++}H^{--}$, with the aim of probing the vertex
$H^{\pm\pm}H^\mp W^\pm$, which is present in the HTM but not in models
with $SU(2)$ singlet scalars.  A cut is imposed on missing energy
(which originates from $H^\pm\to \ell^\pm\nu$) in order to remove the
contribution from $pp\to H^{++}H^{--}$.  Therefore the approach of
Ref.~\cite{Perez:2008ha} contrasts with that of
\cite{delAguila:2008cj} (and our approach), where in the latter the
cuts are designed to keep signal events from both $pp\to H^{++}H^{--}$
and $pp\to H^{\pm\pm}H^\mp$ in order to optimize sensitivity to
$M_{H^{\pm\pm}}$ for a given integrated luminosity.

\noindent The main features of our analysis are :
\begin{itemize}
\item{} In order to analyze the signature of $\ge 3\ell$ as mentioned
above we have to consider the $H^{\pm\pm} H^\mp W^\pm$ vertex. This vertex was not available in \pythia \cite{Sjostrand:2006za} for the HTM model.  For our analysis we have used \calchep \cite{Pukhov:2004ca} and incorporated this vertex in the model file.
\item{} We have included K-factors for both signal and background events. 
\item{} We have performed a detailed realistic detector simulation using the fast detector simulator \atlfast for both signal and background processes.
\item{} We have considered the $\ge 3 \ell$ signature and compared its
  discovery potential with that for the $4 \ell$ signature at
  the LHC.
\end{itemize}

Hereafter, we will refer to electrons and/or muons collectively as
``leptons'' ($\ell = e, \mu$).  We will further assume the idealized
case of 
BR($H^{\pm\pm} \to \ell^\pm \ell^\pm) = 100\%$ and BR($H^{\pm} \to \ell^\pm \nu_\ell) = 100\%$, {\it i.e.}, the decays of the charged Higgs bosons are saturated by the electronic and muonic modes. We do this in order to provide a simple comparison of the discovery potential of the two signatures under investigation. Moreover, such extreme branching ratios are generally used when deriving limits on $M_{H^{\pm\pm}}$ from direct searches. In contrast, we note that representative branching ratios in the HTM were used in \cite{delAguila:2008cj}, for which decay modes of $H^{\pm\pm}$ involving $\tau$ were sizeable. Careful attention was given to secondary electrons and muons which originate from decays like $H^{\pm\pm}\to \mu^\pm\tau^\pm$ followed by $\tau\to \ell\nu\nu$, and their effect on the dilepton invariant mass distribution was studied.  In our analysis the decay modes of $H^{\pm\pm}$ involving $\tau$ are absent, and so there are no such secondary leptons.  For the cases of BR($H^{\pm\pm} \to \ell^\pm \ell^\pm) < 100\%$ and BR($H^{\pm} \to \ell^\pm \nu_\ell) < 100\%$ (as discussed in Section~\ref{section:2} for the HTM) our results will need to be scaled by multiplicative factors of branching ratios.  Moreover, for non-zero BRs of $H^{\pm\pm}$ and $H^\pm$ into final states which contain $\tau$ leptons, the influence of the secondary leptons (which originate from the decay of the $\tau$ leptons) on the signal will need to be included.  For definiteness, we take $M_{H^{\pm\pm}} = M_{H^{\pm}}$ as the mass difference is fairly small for most of the parameter space in the HTM.

\FIGURE{
\epsfig{file=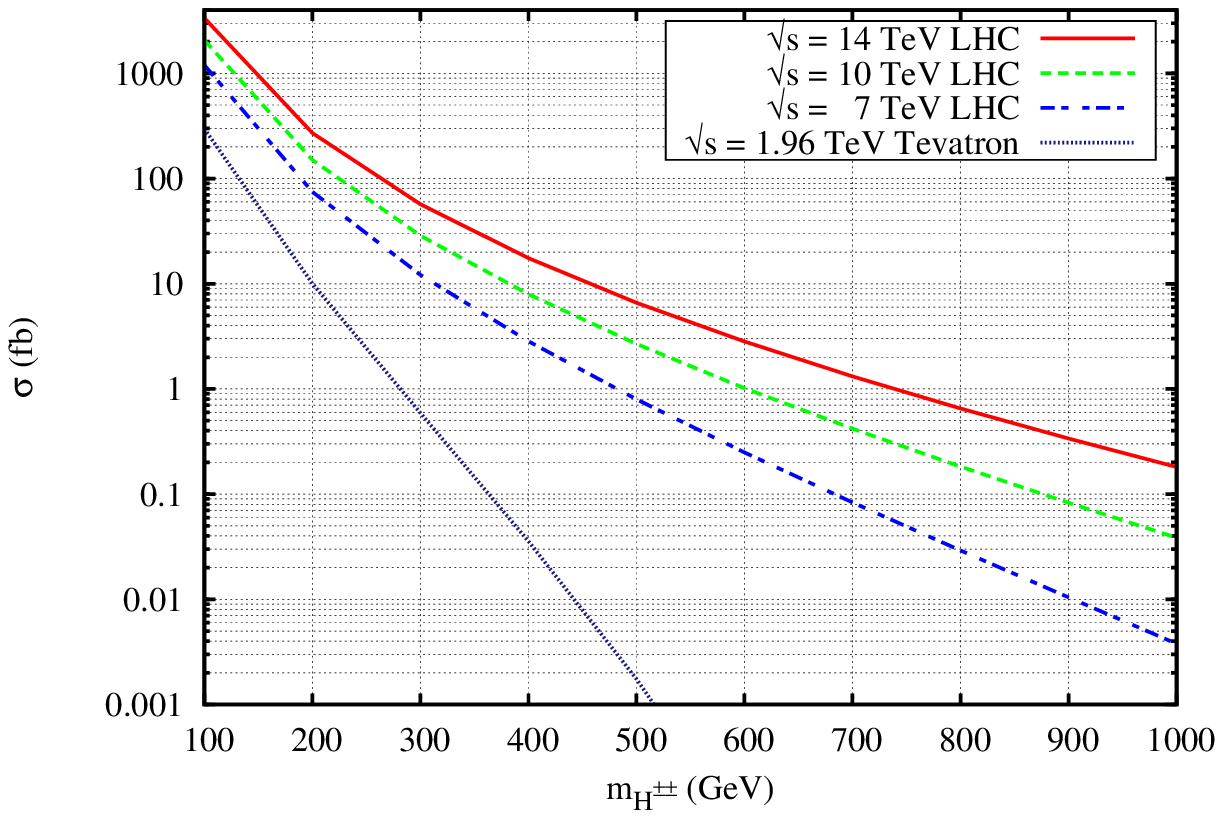,width=.85\textwidth}
\caption{Cross section of inclusive doubly charged Higgs bosons
production (Eq.~\ref{single_prod}) as a function of $M_{H^{\pm \pm}}$.
The K-factor of the processes is taken to be 1.25 for LHC and 1.3 for
Tevatron.} 
\label{fig:1}
}

The cross section for the inclusive production of doubly charged Higgs
bosons at hadronic colliders is plotted in Figure~\ref{fig:1} as a
function of the doubly charged Higgs mass.  In this plot we show the
production cross sections as given in Eq.~(\ref{single_prod}) for the
LHC at the center-of-mass (CM) energy of $\sqrt{s} = 7, 10, 14$ TeV
and for the Tevatron $\sqrt{s} = 1.96$ TeV.  The LHC is expected to
take around 1 fb$^{-1}$ of integrated luminosity at $\sqrt{s} = 7$ TeV
in its first two years of operation.  Subsequently, the machine is
planned to run at the design energy of $\sqrt{s} = 14$ TeV.  For
completeness, we also show the cross section at an intermediate energy
of $\sqrt{s} = 10$ TeV because operation at this energy has been
discussed.  In our later simulations, we have assumed the CM energy of
$\sqrt{s} = 14$ TeV at the LHC.  We have used the leading-order (LO)
CTEQ6L parton distribution functions (PDF) with two-loop $\alpha_s$
running, and identified both the factorization scale $\mu_f$ and the
renormalization scale $\mu_r$ with the partonic CM energy $\hat{s}$. 


\subsection{Framework for event generation \label{subsec:4_1}}

The SM background processes we have considered 
for both the $\ge 3\ell$ and $4\ell$ channels are:   
\begin{itemize}
\item{} $Z Z$ with each of the $Z$'s decaying leptonically. 
\item{} $W^\pm Z$ with each of the weak gauge bosons decaying leptonically. 
\item{} $t \bar{t}$ with $t \to W b$, and $W$'s and $b$ decaying (semi)leptonically. 
\item{} $Z b b$ with $Z$ and $b$ decaying (semi)leptonically. 
\item{} $W b b$ with $W$ and $b$ decaying (semi)leptonically.
\item{} $Z t t$ with $Z$ and $t$ decaying (semi)leptonically.
\item{} $W t t$ with $W$ and $t$ decaying (semi)leptonically.
\item{} $W^\pm W^\mp W^\pm$ with each of the $W$'s decaying leptonically. 
\end{itemize}
%
%
\TABLE[h]{
\caption{Number of signal events generated where K stands for
$10^3$. These events are then scaled to the luminosity of ${\cal L}$ =
10 fb$^{-1}$.}
\begin{tabular}{|c|c|c|c|c|c|c|c|} \hline
$m_{H^{\pm\pm}}$ (GeV) & 200 & 300 & 400 & 500 &
600 & 700 & 800 \\ \hline
Events generated  &  300 K & 150 K & 100 K & 100 K & 80 K & 80 K & 60
K \\ \hline 
\end{tabular}
\label{signal-events}
}

The setup for signal and background event generations is the following:
\begin{itemize}
\item\underline{\bf Signal event generation:} We have used \calchep
v.2.5.4 \cite{Pukhov:2004ca} for calculating cross sections.  For
this purpose we have implemented the relevant interaction vertices in
the \calchep model files.  The partonic level signal events have been
generated using \calchep and then passed to \pythia v.6.4.21
\cite{Sjostrand:2006za} via Les Houches Event (LHE) interface
\cite{Alwall:2007mw} in order to include initial state radiation/final
state radiation (ISR/FSR) effects. 
The number of signal events generated are given in Table~\ref{signal-events}. These events were then scaled to the luminosity
of ${\cal L}$ = 10 fb$^{-1}$. 

\item\underline{\bf Background event generation:} We have generated
the $t \bar{t}$, $ZZ$, $W^\pm Z$ events directly using \pythia, and
the $Zbb$, $W^\pm b b$, $Ztt$, $W^\pm t t$ events first using \calchep
and then interfaced with \pythia for ISR/FSR. 
The number of background events generated are given in Table~\ref{background-events}. These events were scaled to the luminosity of
${\cal L}$ = 10 fb$^{-1}$. 
\end{itemize}

%
\TABLE[h]{
\caption{Number of background events generated where K stands for
$10^3$ and M stands for $10^6$. The second row corresponds to the decay
modes considered, and SL indicates ``semi-leptonic.''  These events are
then scaled to the luminosity of ${\cal L}$ = 10 fb$^{-1}$.}   
\begin{tabular}{|c|c|c|c|c|c|c|c|c|} \hline
Process & ZZ  & WZ & WWW & $t \bar{t}$ & Z bb & W bb & Ztt & Wtt \\ \hline 
Decay modes  & all & all & all & SL & SL & SL & SL & SL \\ \hline 
Events  & 1.5 M & 1.5 M & 300 K & 90 M  & 1.2 M  & 900 K & 150 K & 800 K    \\ \hline 
\end{tabular}
\label{background-events}
}

In order to make more realistic estimates of the signal and background
events, we have further processed both of them through the fast ATLAS
detector simulator \atlfast \cite{atlfast}.  The resulting events have
been analyzed within the ROOT framework. The detector simulator
\atlfast provides simple detector simulation and jet reconstruction
using a simple cone algorithm.  It further identifies isolated
leptons, photons, $b$ and $\tau$ jets, and also reconstructs missing
energy.  In our analysis for the LHC, we have assumed the CM energy of
$14$ TeV and luminosity ${\cal L} = 10$ fb$^{-1}$.  The K-factors for
signals and backgrounds that we have used are listed in
Table~\ref{table:k-fac}.

%
\TABLE[h]{
\caption{\sl K-factors for the background and signal processes at LHC}
\begin{tabular}{|c|c|c|c|c|c|c|c|} \hline
Process & $WZ$ & $ZZ$ & $t \bar{t}$ & $Z b b$ & $Z t t$ & $W
b b$ & $H^{\pm\pm} H^{\mp\mp}, H^{\pm\pm} H^\mp$ \\  \hline
K-factors & 1.5 & 1.35 & 1.67 & 2.4 & 1.35 & 2.57 & 1.25 \\ \hline
\end{tabular}
\label{table:k-fac}
}

In order to improve the search potential for the doubly charged Higgs
boson at the LHC, we want to advocate the D0 search strategy
\cite{Abazov:2008iy} of looking for $\ge 3 \ell$.  This is in contrast
with the current LHC search strategy \cite{Hektor:2007uu} of looking
for exactly four leptons.  Accordingly, we will present our simulation
results for exactly four-lepton signature and $\ge 3$ lepton
signature. 


\subsection{Signature of four leptons
\label{subsec:4_2}} 

This is the signature ($4\ell$, $\ell=e,\mu$) when we have the pair
production of doubly charged Higgs bosons via the process $p p \to
H^{\pm\pm} H^{\mp\mp}$ followed by leptonic decays $H^{\pm \pm} \to
\ell^\pm \ell^\pm$.  This signature has been simulated within the
context of LHC in
Refs.~\cite{Hektor:2007uu,Rommerskirchen:2007jv,delAguila:2008cj,Azuelos:2005uc}.

\noindent We use the following {\it pre-selection cuts} on signal and
background events \cite{delAguila:2008cj} :  
\begin{itemize}
\item{} There are exactly four leptons with two for each charge sign
($\ell^+\ell^+ \ell^-\ell^-$) in each event. 
\item{} Each of the leptons has $|p_T^\ell| > 5$ GeV and
pseudorapidity in the range $|\eta| < 2.5$. 
\item{} Amongst the four leptons, at least two of the leptons have
$|p_T^\ell| > 30$ GeV.  This cut reduces the backgrounds where the
leptons originate from the semileptonic $b$ decays as they tend to be
less energetic. 
\item{} Opposite-sign dilepton invariant mass cut: $m_{\ell^+ 
\ell^-} > 20$ GeV.  This is done in order to suppress the backgrounds
where the opposite-sign lepton pair comes from a photon. 
\end{itemize}

\noindent We impose additional cuts to further improve the signal
significance:  
\begin{itemize}
\item[(a)]
The $Z$ window cut.  The invariant mass of opposite-sign dileptons is
required to be sufficiently far from the $Z$ mass: $|m_{\ell^\pm
\ell^\mp} - M_Z | > 10$ GeV.   This removes events where the leptons
come from the $Z$ decay. 
\item[(b)]
The $H_T$ cut.  One can also use the total transverse energy ($H_T$) as
a parameter to distinguish signals from backgrounds.  The total
transverse energy is defined as 
\begin{equation}
H_T = \sum_{\ell, ~ \et} |\vec{p}_T| ~.
\end{equation}
The $H_T$ distribution of signals tends to peak around the heavy
particle mass, and hence this cut can be used to suppress the SM
background processes that involve relatively light particles.  We will
show the results for $H_T > 300$ and $ 500$ GeV.  We note that this
$H_T$ cut was also used in the study of the $4\ell$ signature in
Ref.~\cite{Hektor:2007uu}, but it was not used in the study of
Ref.~\cite{delAguila:2008cj}. 
\end{itemize}

%
\TABLE[ht]{
\caption{ Background and signal events surviving the cuts for 
  \underline{exactly 4-lepton} final states. For these numbers we have
  taken ${\cal L} = 10$ fb$^{-1}$ and $\sqrt{s} = 14$ TeV.} 
\begin{tabular}{|c|c|c|c|c|c|c|c|c|} \hline 
 & \multicolumn{6}{c|}{Backgrounds}  & \multicolumn{2}{c|}{Signal ($M_{H^{\pm\pm}}$)} \\ \hline 
Cut & $WZ$ & $ZZ$ & $t \bar{t}$ & $Z bb$ & $Z tt$ &  $W tt$ & 200 GeV & 600 GeV \\  \hline 
Pre-selection & 0.2 & 130.5 & 1.3 & 0.2 & 122.6 & 0.1 & 400.1 & 4.2 \\ \hline 
$|m_{\ell^+ \ell^-} - m_Z| > 10$ GeV & 0.1 & 2.1 & 0.3 & 0 & 2.1 & 0.1 & 330.6 & 4.1 \\ \hline
$H_T > 300$ GeV & 0 & 0.4 & 0 & 0 & 1.2 & 0 & 327.9 & 4.1  \\ \hline
$H_T > 500$ GeV & 0 & 0.1 & 0 & 0 & 0.3 & 0 & 222.9 & 4.1  \\ \hline
${\cal S}$      &   & &  & &   &            & 48.7  & 3.7  \\ \hline 
\end{tabular}
\label{table:4lepton}
}

The effects of the cuts on signal and background events for two indicative
masses of the doubly charged Higgs boson are given in
Table~\ref{table:4lepton}.  After the pre-selection cuts, the dominant
backgrounds are $ZZ$ and $Ztt$, which are then significantly reduced
by the $Z$ window cut.  For large $M_{H^{\pm\pm}}$, the $Z$ window and
$H_T$ cuts only reduce the number of signal events by a negligible
amount and hence the significance of the signal is improved.


\subsection{Signature of at least three leptons \label{subsec_3}} 

\FIGURE[t]{
\epsfig{file=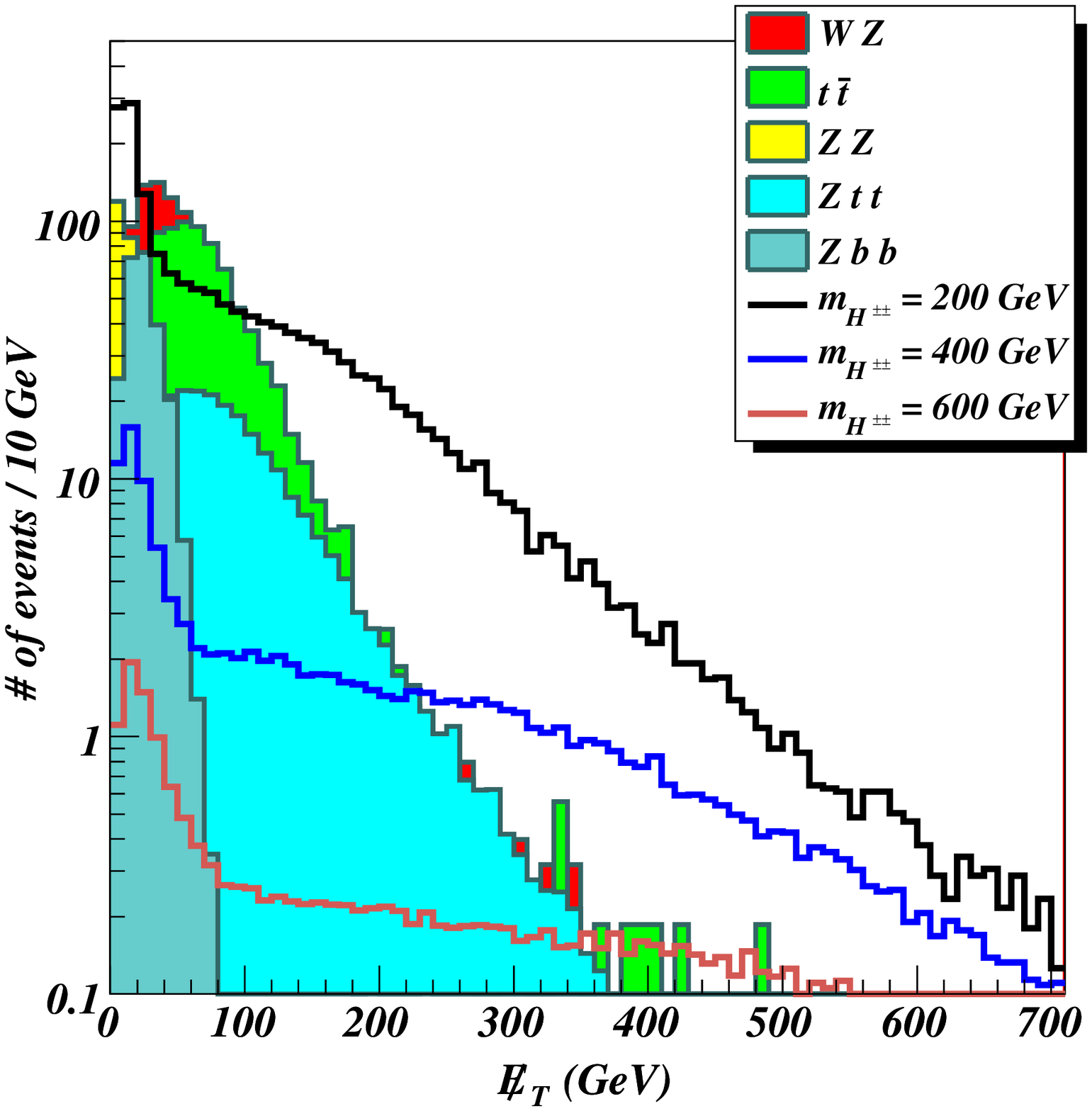,width=0.5\textwidth}
\caption{$\et$ distribution for doubly charged Higgs production
(signal) and SM backgrounds after imposition of the pre-selection
cuts.  We have taken the CM energy of $\sqrt{s} = 14$ TeV and
luminosity ${\cal L} = 10$ fb$^{-1}$.}  
\label{fig:3l:misset}
}

The study of the signature with exactly three leptons ($3\ell$) was done in Ref.~\cite{delAguila:2008cj}, with the motivation of comparing the detection prospects of three distinct types of seesaw-based models of neutrino mass generation (one of which being the HTM).  It was acknowledged that such a signature is not necessarily the one which optimizes the discovery potential in a given model.  In this paper, we are concerned with optimizing the sensitivity to $M_{H^{\pm\pm}}$ and thus we consider the signature of $\ge 3 \ell$ as done in the D0 search \cite{Abazov:2008iy}.  As emphasized earlier, both the pair production and single production of the doubly charged Higgs boson contribute to the signature and hence increase the signal events.  For this analysis, we have used cuts similar to those in Section~\ref{subsec:4_2}.  The only difference is that instead of singling out events with exactly four leptons, here we select events that have at least three leptons, of which two have the same sign.  We would like to note that Ref.~\cite{delAguila:2008cj} used a slightly different set of {\it pre-selection} cuts for studying the {\it exactly three-lepton} signature.  For the {\it exactly three-lepton} signature, they chose events that had {\it at least} two {\it same-sign} leptons with $p_T > 30$ GeV.  Also, the {\it pre-selection} cuts used in Ref.~\cite{delAguila:2008cj} were slightly different for exactly four-lepton and exactly three-lepton signatures, whereas in order to make a comparison between the four-lepton and $\geq$ three-lepton signatures we are using the same set of cuts for both.

The effects of the cuts on signal and background events for the $\ge 3
\ell$ signature are given in Table~\ref{table:ge3lepton}. It is
evident that the discovery reach of the current signature is better
than the $4\ell$ signature. 

Also shown in both Tables~\ref{table:4lepton} and
\ref{table:ge3lepton} are the significance.  As signal and background
events become fewer after the cuts, it is necessary to employ Poisson
statistics to estimate the significance of the signal.  We use the
significance estimator \cite{Ball:2007zza} 
\begin{equation}
{\cal S} = \sqrt{2 \left\{ n_0 \ln\left(1 + \frac{s}{b}\right) - s \right\}},
\end{equation}
where $b$ is the expected number of background events and $n_0$ is the
number of observed events.  Accordingly, the number of signal events
is $s = n_0 - b$.  This estimator is based on a log-likelihood ratio,
and follows very closely the Poisson significance.  In the limit of
$s/b \ll 1$, it reduces to the simple estimator $s/\sqrt{b}$. 

%
\TABLE[ht]{
\caption{Background and Signal events surviving the cuts for
  \underline{at least 3 leptons} in the final state. We have taken ${\cal L}
  = 10$ fb$^{-1}$ and $\sqrt{s} = 14$ TeV.}
\begin{tabular}{|c|c|c|c|c|c|c|c|c|c|} \hline
 & \multicolumn{7}{c|}{Backgrounds} & \multicolumn{2}{c|}{Signal
($M_{H^{\pm\pm}}$)} \\ \hline 
Cuts $\Downarrow$ & $WZ$ & $WWW$ & $ZZ$ & $t \bar{t}$ & $Z bb$
& $Z tt$ & $W tt$ & 200 & 600  \\ \hline 
Pre-selection & 591.7 & 3.5 & 203.6 & 159.9 & 57.7 & 212.5 & 9.7
& 1570.4 & 17.6 \\ \hline
$|m_{\ell^+ \ell^-} - m_Z| > 10$ GeV & 50.9 & 2.7 & 12.1 & 113.2 & 0.9
& 33.4 & 7.4 & 1397.8 & 17.3 \\ \hline
$H_T > 300$ GeV & 7.5 & 1.1 & 1.6 & 8.9 & 0   & 17  & 3.4 & 1351.1 & 17.3 \\ \hline
$H_T > 500$ GeV & 1.7 & 0.3 & 0.4 & 0.9 & 0   & 3.2 & 0.6 & 796.2 &
17.3  \\ \hline 
${\cal S}$   &     &     &     &     &     &     &     &  77.4 &
5 \\ \hline
\end{tabular}
\label{table:ge3lepton}
}

\FIGURE[t]{
\epsfig{file=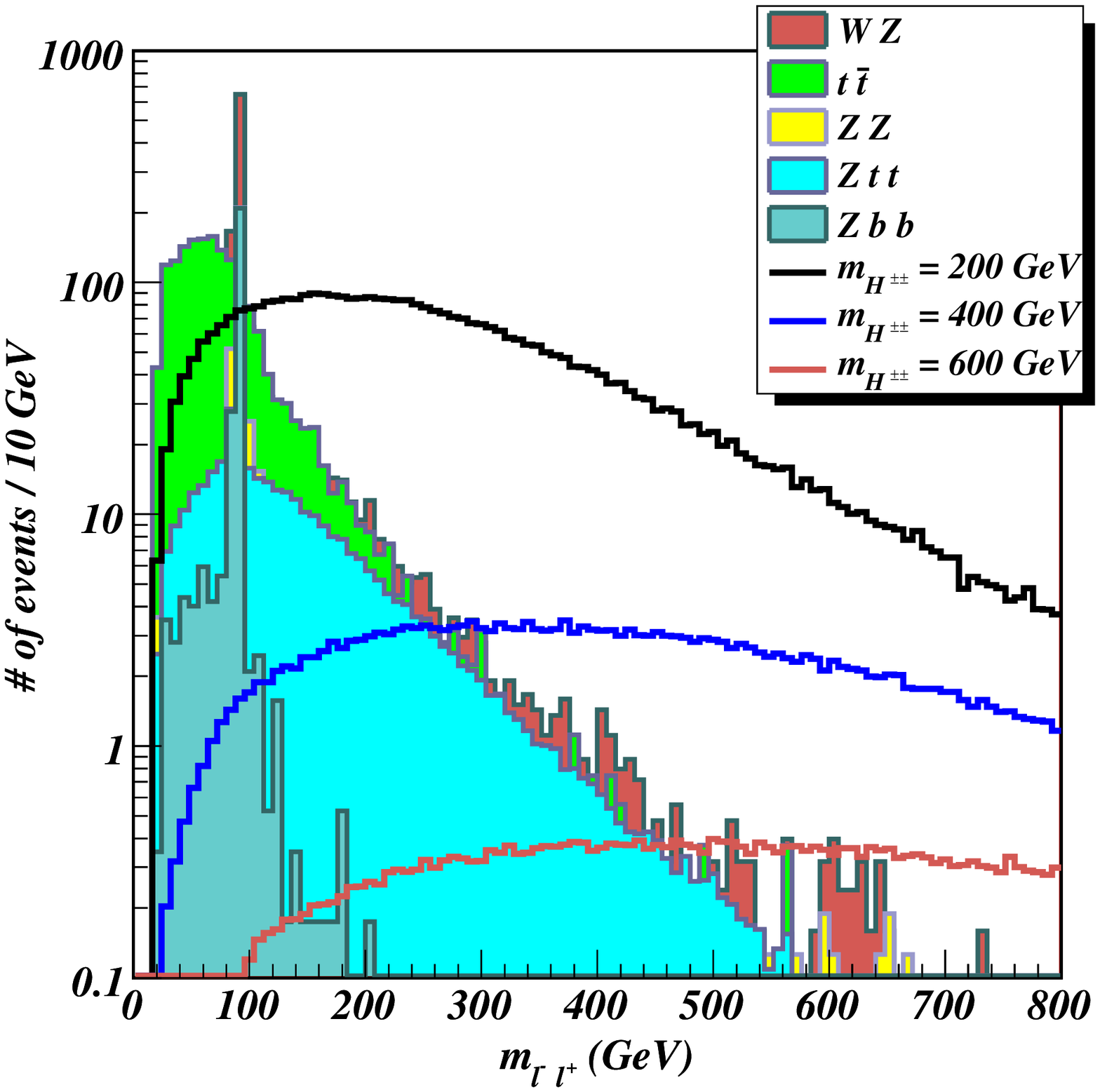,width=0.5\textwidth}\hspace*{.4cm}
\epsfig{file=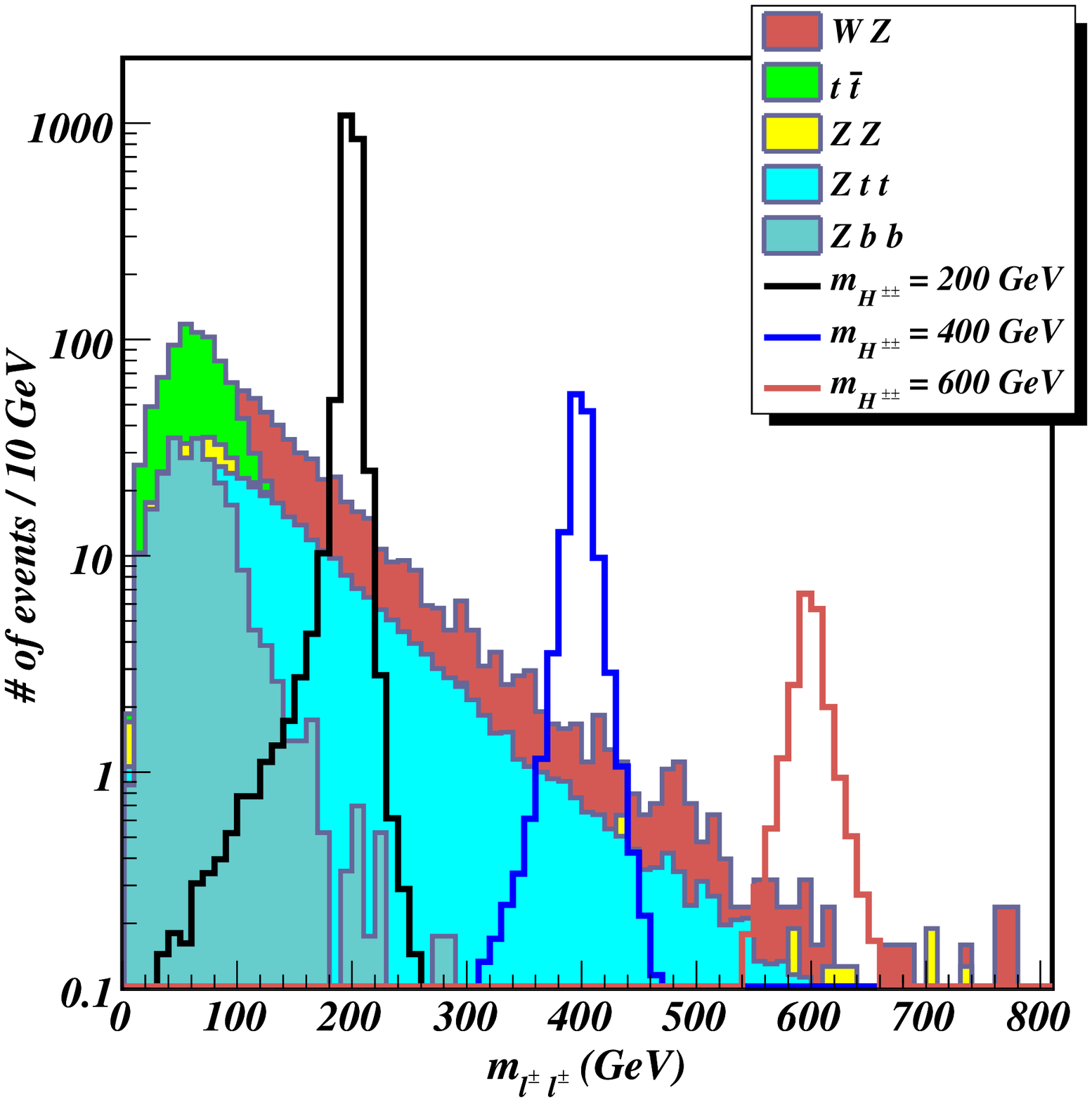,width=0.5\textwidth}
\caption{(a) Left panel: opposite-sign dilepton invariant mass
distribution, (b) Right panel: same-sign dilepton invariant mass
distribution, for both signal and background events.}
\label{fig:3l:leptoninv}
}

\FIGURE[h]{
\epsfig{file=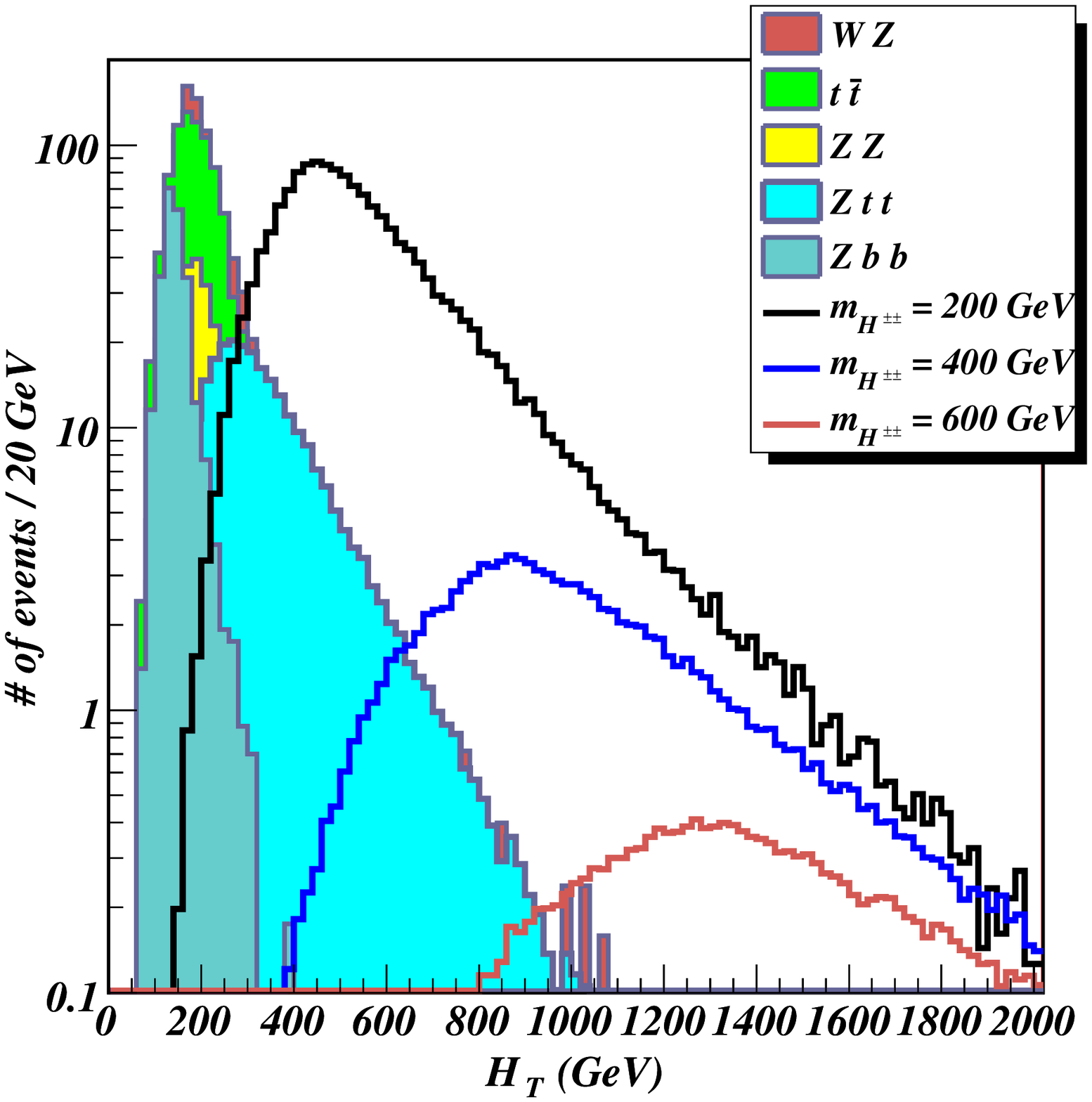,width=0.6\textwidth}
\caption{ Total transverse energy ($H_T$) distribution for signal
and SM background events. }
\label{fig:3l:ht}
}

The distributions presented in Figures~\ref{fig:3l:misset},
\ref{fig:3l:leptoninv}, and \ref{fig:3l:ht} are for the luminosity of
${\cal L} = 10$ fb$^{-1}$ and CM energy of $14$ TeV at the LHC.  In
Figure~\ref{fig:3l:misset}, we show the $\et$ distribution for both
signal and background events.  The background events concentrate at
lower $\et$, whereas the signal events extend to the higher region as
they contain more energetic neutrinos coming from the singly charged
Higgs bosons.  In Figure~\ref{fig:3l:leptoninv}, we show two dilepton
invariant mass distributions.  As can be seen in the opposite-sign
dilepton invariant mass distribution (left panel), the event
distribution tends to peak around the $Z$ mass as the lepton pairs
mostly originate from the decay of the $Z$ boson.  These backgrounds
can be readily reduced by imposing the $Z$ window cut on the
opposite-sign dilepton invariant mass.  On the other hand, the
same-sign dilepton invariant mass distribution peaks around the doubly
charged Higgs mass whereas the SM processes form a continuous
background, as shown in the right panel of
Figure~\ref{fig:3l:leptoninv}.

Finally we show the total transverse energy distribution in Figure~\ref{fig:3l:ht}.  The $H_T$ distribution peaks around the total mass of the heavy particles produced in the hard process.  Hence this variable serves as a very useful discriminator between the signal and SM backgrounds, especially for relatively heavy doubly charged Higgs masses.  We emphasize that we have performed the first study of the dependence of the three-lepton signature on the parameter $H_T$.  It is evident that a clear signal for $H^{\pm\pm}$ can be obtained by using a cut on $H_T$ to reduce backgrounds, even before plotting the dilepton invariant mass distribution (the latter providing information on $M_{H^{\pm\pm}}$).

\FIGURE[ht]{
\epsfig{file=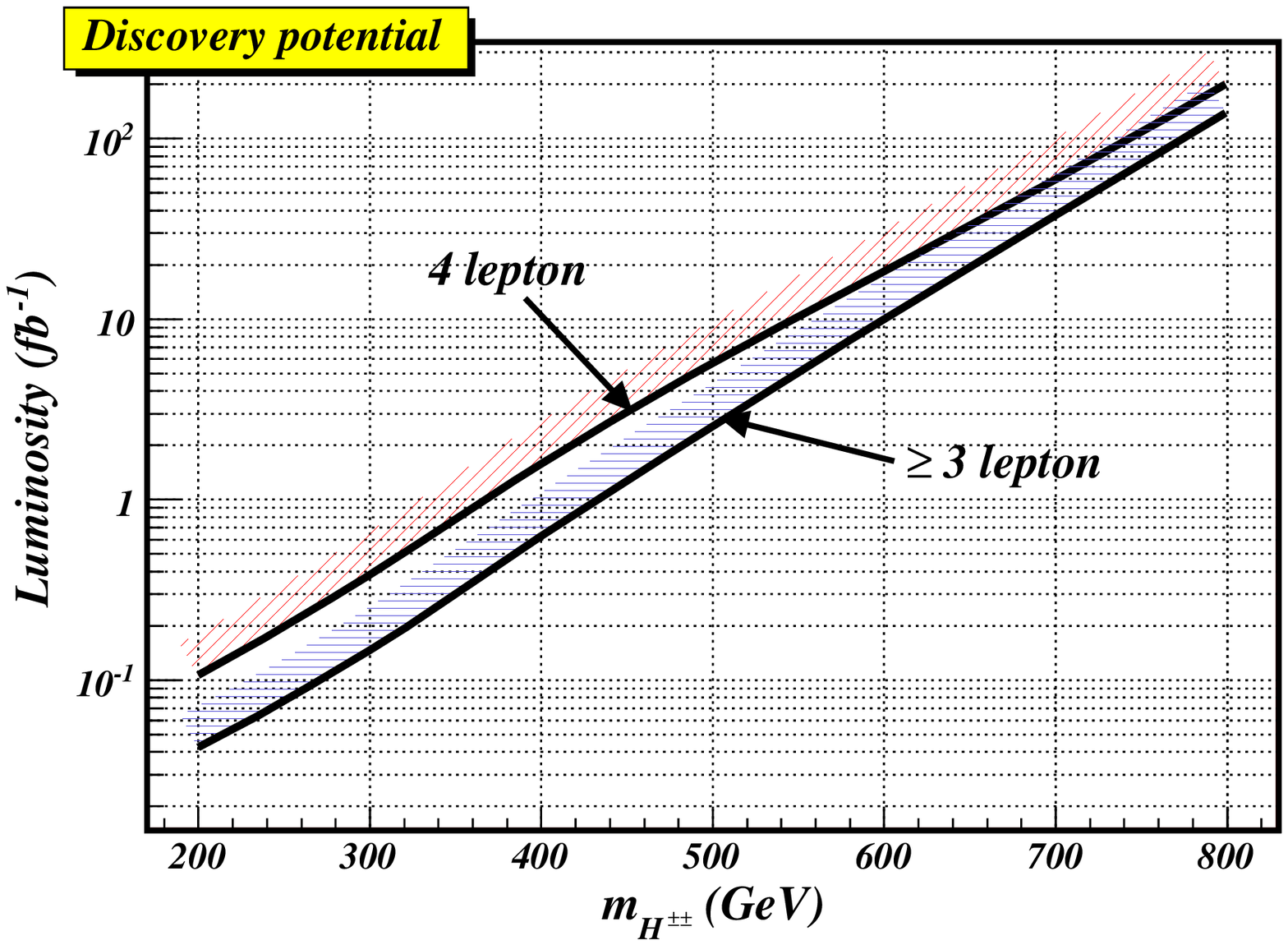,width=.8\textwidth}
\caption{Luminosity required for a $5 \sigma$ discovery of the
doubly charged Higgs boson as a function of its mass at the LHC.  The
two curves correspond to the exactly 4-lepton signature and the $\ge
3\ell$ signature, respectively. } 
\label{fig:discovery}
}


\section{Discussions and Summary \label{section:5}}

In this paper we have analyzed the leptonic signatures of the
production of a doubly charged Higgs boson at LHC.  For this purpose
we have used \calchep to generate the signal events, and then
interfaced it with \pythia.  For more realistic estimates of signal
and background events, we have used fast ATLAS detector simulator
\atlfast.  We have also included relevant K-factors for both signal
and backgrounds in our analysis. 

Using the significance estimator, we have estimated the LHC discovery
potential for the doubly charged Higgs boson in the 4-lepton mode from
the pair production only and the $\ge 3$ lepton mode from the
inclusive production, {\it i.e.}, both $pp\to H^{++}H^{--}$ and $pp\to
H^{\pm\pm}H^{\mp}$.  We have performed the first simulation of the
$\ge 3$ lepton channel.  Our result is shown in
Figure~\ref{fig:discovery}, where the luminosity required to make a $5
\sigma$ discovery of the doubly charged Higgs boson is plotted as a
function of its mass. It is clear that the discovery potential for
$H^{\pm\pm}$ at the LHC through the $\ge 3 \ell$ mode is significantly
better than the 4-lepton mode.  For an integrated luminosity of $10$
fb$^{-1}$, for example, one detector at LHC alone can reach $\sim 600$
GeV for the former and $\sim 550$ GeV for the latter. 

In our analysis of multi-lepton signatures we have not considered the QCD background where a jet is misidentified as a lepton.  Although the probability of misidentification is quite small \cite{Azuelos:2005uc}, the QCD production cross-sections are many orders of magnitude larger than the multi-lepton ($3\ell$ and $4\ell$) cross-sections and hence can contribute to the backgrounds.  Importantly, the QCD background for the four-lepton signal will be smaller than that for the three-lepton signal.  Hence the inclusion of the QCD background would introduce a systematic error into the estimates given in Figure~\ref{fig:discovery}, which could alter the significance of our results.  In an actual experiment, the QCD background can be estimated from the data.  For a more realistic study, one should do a full detector simulation, optimize the cuts, and take into account the statistical and systematic uncertainties. We hope that this study will motivate our experimental colleagues at hadronic colliders to update their analyses by considering the following points:
\begin{itemize}
\item{\bf Tevatron:} Include the process $p \bar{p} \to H^{\pm \pm}
H^\mp$ in the analysis when searching for the $\ge 3 \ell$ signature. 
\item{\bf LHC:} Search for the $\ge 3 \ell$ signature to increase the
LHC discovery reach of the doubly charged Higgs boson.  
\end{itemize}


\acknowledgments

C.~C. would like to thank Neil Christensen for valuable help with the
\calchep package.  N.~G. would like to thank Michael Klasen and Arun
Kumar Nayak for their comments/suggestions, and acknowledge the
support from NCTS, Taiwan for his visit to NCU, where this work was
initiated.  This work of C.~C. was supported in part by the Grant
No.~NSC~97-2112-M-008-002-MY3 and NCTS.  A.~G.~A was supported by the
``National Central University Plan to Develop First-class Universities
and Top-level Research Centers'', and 
by a Marie Curie Incoming International Fellowship,
FP7-PEOPLE-2009-IIF, Contract No. 252263.   
 N.~G. was supported by a grant from University Grants Commission
(UGC), India under project number 38-58/2009 (SR).


\end{document}